\documentclass[usenatbib]{mn2e}
\usepackage{graphicx}

\def\apj{\rm ApJ}
\def\apjl{\rm ApJL}
\def\apjs{\rm ApJS}
\def\aj{\rm AJ}
\def\mnras{\rm MNRAS}
\def\nat{\rm Nature}

\def\aap{\rm AAP}

\def\gax{\mathrel{\raise.3ex\hbox{$>$}\mkern-14mu\lower0.6ex\hbox{$\sim$}}}
\def\lax{\mathrel{\raise.3ex\hbox{$<$}\mkern-14mu\lower0.6ex\hbox{$\sim$}}}
\def\gtorder{\mathrel{\raise.3ex\hbox{$>$}\mkern-14mu
             \lower0.6ex\hbox{$\sim$}}}
\def\ltorder{\mathrel{\raise.3ex\hbox{$<$}\mkern-14mu
             \lower0.6ex\hbox{$\sim$}}}

\begin{document}

\title [TDE Abundance Anomalies]
  {Abundance Anomalies In Tidal Disruption Events}

\author[C.~S. Kochanek]{ C.~S. Kochanek$^{1,2}$\\
  $^{1}$ Department of Astronomy, The Ohio State University, 140 West 18th Avenue, Columbus OH 43210 \\
  $^{2}$ Center for Cosmology and AstroParticle Physics, The Ohio State University,
    191 W. Woodruff Avenue, Columbus OH 43210 
   }

\maketitle

\begin{abstract}
The $\sim 10\%$ of tidal disruption events (TDEs) due to stars more massive 
than $M_*\gtorder M_\odot$ 
should show abundance anomalies due to stellar evolution in helium, carbon and nitrogen, but 
not oxygen.  Helium is always enhanced, but
only by up to $\sim 25\%$ on average because it becomes inaccessible
once it is sequestered in the high density core as the star leaves
the main sequence.  However, portions of the debris associated with the
disrupted core of a main sequence star can be enhanced in helium by factors of 
$2$-$3$ for debris at a common orbital period.  These helium abundance variations may be a 
contributor to the observed diversity of hydrogen and helium line 
strengths in TDEs.   A still more striking anomaly is the
rapid enhancement of nitrogen and the depletion of carbon due to 
the CNO cycle -- stars with $M_* \gtorder M_\odot$ quickly show an increase
in their average $N/C$ ratio by factors of $3$-$10$.
Because low mass stars evolve slowly and high mass stars are rare, TDEs 
showing high $N/C$ will
almost all be due to $1$-$2M_\odot$ stars disrupted on the main
sequence.  Like helium, portions of
the debris will show still larger changes in C and N, and the
anomalies decline as the star leaves the main sequence.  
The enhanced $\left[ N/C \right]$ abundance ratio of these TDEs provides
the first natural explanation for the rare, nitrogen rich quasars
and also explains the strong nitrogen emission seen in ultraviolet
spectra of ASASSN-14li.
\end{abstract}


\section{Introduction}
\label{sec:introduction}

If a star passes too close to a supermassive black hole, it
can be completely or partially destroyed by tides.  Portions
of the debris are then accreted by the black hole, leading
to a luminous tidal disruption event (TDE, e.g., \citealt{Lacy1982}, 
\citealt{Rees1988}, \citealt{Evans1989}).  
\cite{Arcavi2014} noted that TDEs show a remarkable diversity
in the relative strengths of the hydrogen H$\alpha$/H$\beta$
Balmer lines and the HeII 4686\AA\ emission line. 
In the most extreme
cases, the broad HeII emission lines are significantly stronger
than the Balmer lines (PS1-10jh, \cite{Gezari2012b}; 
PTF09ge, \cite{Kasliwal2009}, ASASSN-15oi REF), while others show both strong HeII
and Balmer emission lines (SDSS~J074820.66+471214.6, \cite{Wang2012};
ASASSN-14ae, \cite{Holoien2014}; ASASSN-14li, \cite{Holoien2015}).
There are also some that appear hydrogen dominated (PTF09djl, PTF09axc, \cite{Arcavi2014};
TDE2, \cite{vanvelzen2011}, and possibly PS1-11af, \cite{Chornock2014}).
This is quite curious because in typical AGN and quasars the HeII 4686\AA\
emission line is generally very weak compared to H$\alpha$
and H$\beta$ (e.g. \citealt{vandenberk2001}).

Debates over the origin of strong HeII emission in TDEs have 
mainly focused
on PS10jh, where \cite{Gezari2012b} proposed that the spectrum,
with strong HeII 4686\AA\ and undetected H Balmer emission ($\gtorder5$:$1$
ratio), could only be explained by disrupting a helium star. 
\cite{Guillochon2014}, based on the quasar broad line region
models of \cite{Korista2004}, argued that there were parameter
ranges that could almost reproduce the observed line ratios
at Solar abundance.  Their physical explanation 
for the origin of the strong helium emission (``matter bounding'' 
a quasar broad line region to include only the inner, higher 
ionization line regions) is not, however, correct. As explained in 
\cite{Gaskell2014}, the key at Solar metallicity lies in finding a density regime
where line opacities can drive the line ratios close to the
thermal limit where the HeII 4686\AA\ 
line is $(6563/4686)^4=3.8 $ times stronger than H$\alpha$.

\cite{Strubbe2015} broadly argue that the underlying {\tt CLOUDY}
models used by \cite{Guillochon2014} and \cite{Gaskell2014} are
inapplicable to the TDE problem.  First, they argue that the 
AGN-like spectra do not match the observed spectral energy 
distributions of TDEs.  It is certainly true that some optical/UV 
TDEs have weak or undetected X-ray emission, but others have
X-ray emission indicating the existence of a hard tail (e.g. \citealt{Holoien2015}).
Second, they argue that an optically thin ionized medium
cannot have a line ratio exceeding the volume emissivity
ratio for a fully ionized He$^{++}$ region,
other than by changing the underlying abundances from
Solar ($n(He)/n(H)\simeq 0.09$ for $X=0.70$, $Y=0.28$ and 
$Z=0.02$).  Third, they argue that the {\tt CLOUDY}
models used by \cite{Guillochon2014} and \cite{Gaskell2014}
assume thermal line widths ($\sim 10$~km/s), far smaller than
the observed line widths ($>10^3$~km/s).  Hence, for the same
physical parameters, the lines are far less optically thick
than assumed by the {\tt CLOUDY} models. \cite{Roth2015b},
however, find that extreme hydrogen to helium line ratios
are achievable at Solar metallicity using broad line widths
and radiation transfer models better suited
to high optical depths than {\tt CLOUDY}.

All these calculations have assumed that the debris has Solar
composition.   This may be true at birth on the zero age
main sequence (ZAMS), but it becomes increasingly less so
as the star evolves.  In particular, the star becomes steadily
more helium rich as it evolves.  Hydrogen burning through the
CNO cycle also
modifies carbon, nitrogen and oxygen abundances,
primarily by reducing the amount of carbon and increasing
the amount of nitrogen.  There are also changes in lithium
and beryllium abundances and CNO isotope ratios, but 
these will not be observable in TDEs because of Doppler
line broadening.

For a normal star, these internal changes have no external consequences
until close to death, when convection on the giant branch
can mix material from the hydrogen burning zone to the surface
(``dredge up'').  In a TDE, however, material processed by 
nuclear reactions is revealed without any need to wait for
these late phases of stellar evolution, and some of the
debris will have abundances and metallicities that are never 
observed in stellar atmospheres (other than Wolf-Rayet stars) 
or the interstellar medium.  In \S2 we examine
the evolution of the helium and CNO abundances for stars with
a range of masses and how this will manifest itself in
a TDE.  In \S3 we discuss nitrogen rich quasars and suggest that
they are likely the remnants of recent TDEs.  
We summarize the results and their
observational implications in \S4.
In particular, we note that the ultraviolet spectrum by
\cite{Cenko2015} of the TDE 
ASASSN-14li (\citealt{Holoien2015}) bears a striking similarity to spectra of nitrogen
rich quasars.  

\begin{figure}
\centerline{\includegraphics[width=3.5in]{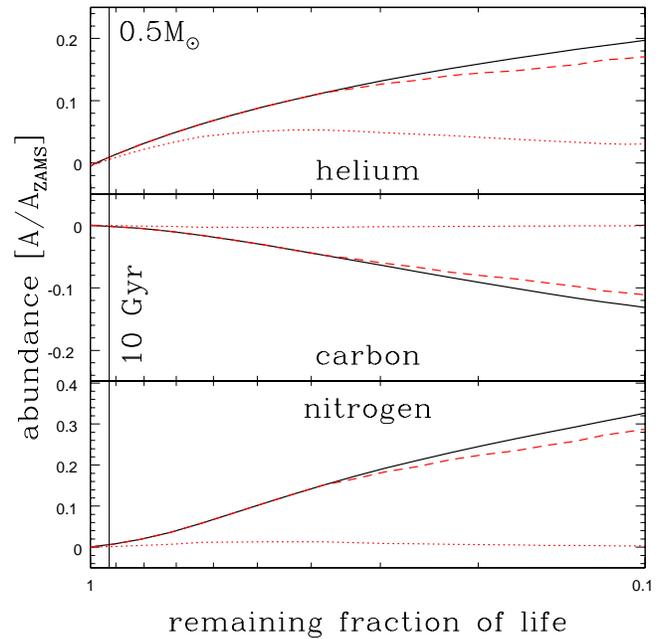}}
\caption{
  Change in abundance $\left[ A(t)/A(0)\right]$ from that on the ZAMS 
  for helium (top), carbon (middle) and nitrogen (bottom) as a function
  of the remaining fractional life time of a $0.5M_\odot$ star.  The heavy solid curve is
  for the entire star and the red dashed (dotted) line is for the material
  outside the inner (outer) edge of the hydrogen burning region (defined
  by $\epsilon > 1$~erg~g$^{-1}$~s$^{-1}$).  Main sequence turn off 
  occurs as the curve for the inner edge of the hydrogen burning region 
  diverges from the results for the entire star.
  A vertical line marks an
  age of $10$~Gyr. Low mass stars, like this $0.5M_\odot$ example, do
  not have time to develop abundance anomalies given the age of the universe. 
  }
\label{fig:mass0.5}
\end{figure}

\begin{figure}
\centerline{\includegraphics[width=3.5in]{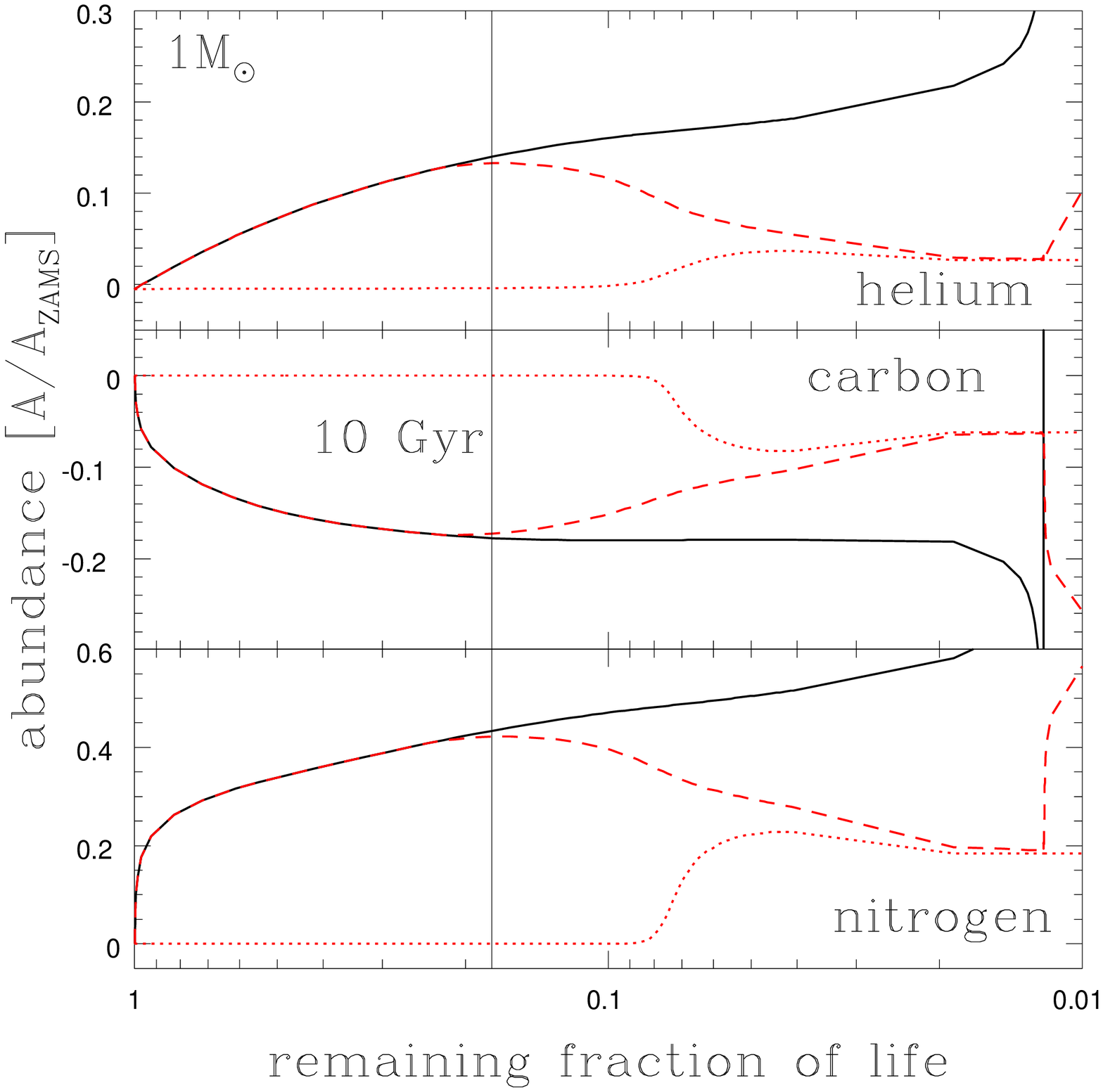}}
\caption{
  Change in abundance $\left[ A(t)/A(0)\right]$ from that on the ZAMS
  for helium (top), carbon (middle) and nitrogen (bottom) as a function
  of the remaining fractional life time of a $1.0M_\odot$ star.  The heavy solid curve is
  for the entire star and the red dashed (dotted) line is for the material
  outside the inner (outer) edge of the hydrogen burning region.
  A vertical line marks an
  age of $10$~Gyr.  For a constant star formation rate over the age of   
  the universe, Sun-like stars will be disrupted at a roughly random evolutionary 
  phase.  For an old stellar population, Sun-like stars will be near or
  past the main sequence turnoff.
  }
\label{fig:mass1.0}
\end{figure}

\begin{figure}
\centerline{\includegraphics[width=3.5in]{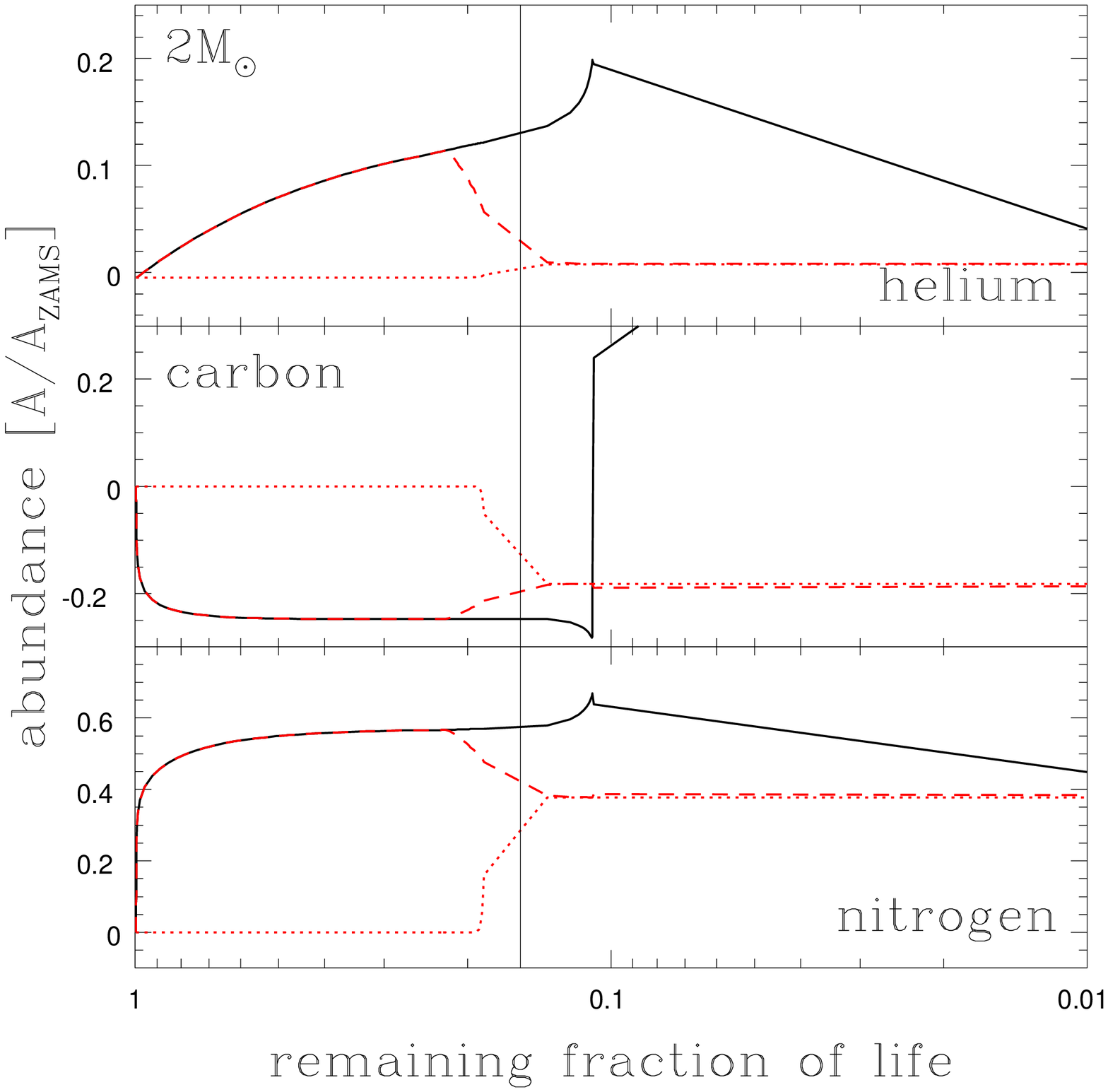}}
\caption{
  Change in abundance $\left[ A(t)/A(0)\right]$ from that on the ZAMS
  for helium (top), carbon (middle) and nitrogen (bottom) as a function
  of the remaining fractional life time of a $2.0M_\odot$ star.  The heavy solid curve is
  for the entire star and the red dashed (dotted) line is for the material
  outside the inner (outer) edge of the hydrogen burning region.
  A vertical line marks an
  age of $1$~Gyr.  For shorter lived massive stars, the evolutionary
  state at the time of disruption will become nearly random and independent
  of the star formation history.
  }
\label{fig:mass2.0}
\end{figure}

\begin{figure}
\centerline{\includegraphics[width=3.5in]{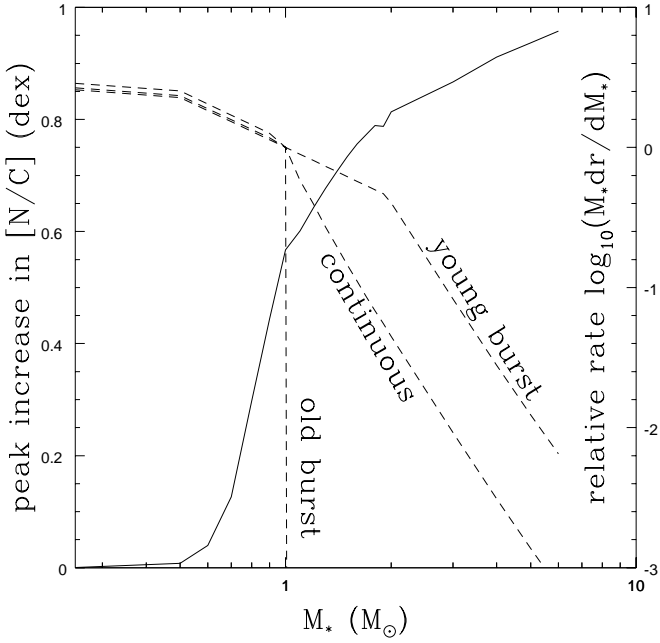}}
\caption{
  Peak enhancement in $\left[ N/C\right]$ relative to the ZAMS abundance
  ratio (left scale, solid curve) as a function of stellar mass $M_*$ assuming a maximum stellar 
  age of 10~Gyr.  Low mass stars ($M_*\ltorder M_\odot$) have no time to 
  develop $\left[ N/C\right]$ anomalies given the age of the universe, 
  while high mass stars quickly generate large anomalies as the CNO
  cycle becomes more important.  The dashed curves (right scale) roughly
  show the scaling of TDE rates per logarithmic mass interval ($ M_* (dr/dM_*)$)
  for an old star burst (2~Gyr of star formation 10~Gyr ago), continuous 
  star formation, and a recent star burst (last 1~Gyr).
  }
\label{fig:cn}
\end{figure}

\begin{figure}
\centerline{\includegraphics[width=3.5in]{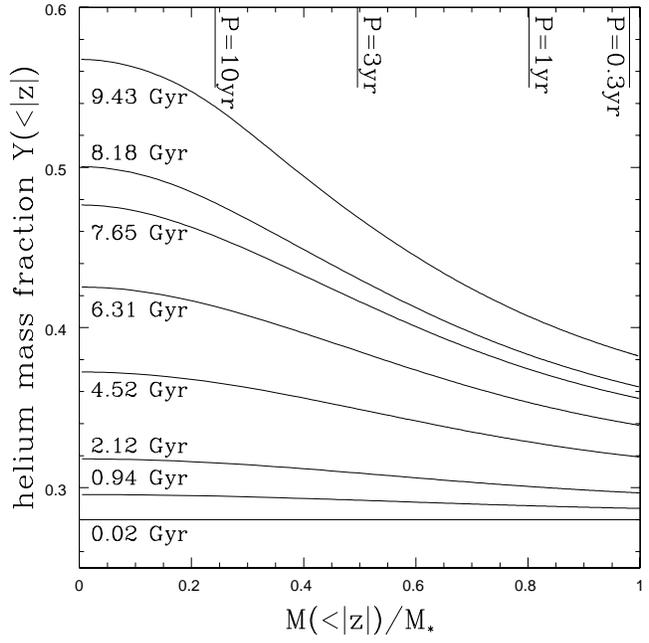}}
\caption{
  Enclosed helium mass fraction $Y(<|z|)$ as function of the enclosed mass 
  fraction $M_*(<|z|)/M_*$ for a $M_*=1M_\odot$ star.  The coordinate $z$
  is the distance from the center of the star towards the black hole
  at pericenter. Profiles are shown
  at the indicated times from $0.02$~Gyr to core hydrogen exhaustion near
  $9.43$~Gyr.  Vertical bars show the orbital period assuming 
  $M_{BH}=10^6 M_\odot$ and a disruption at $R_p=R_*(M_{BH}/M_*)^{1/3}$
  at roughly the age of the Sun.
  }
\label{fig:helium}
\end{figure}

\begin{figure}
\centerline{\includegraphics[width=3.5in]{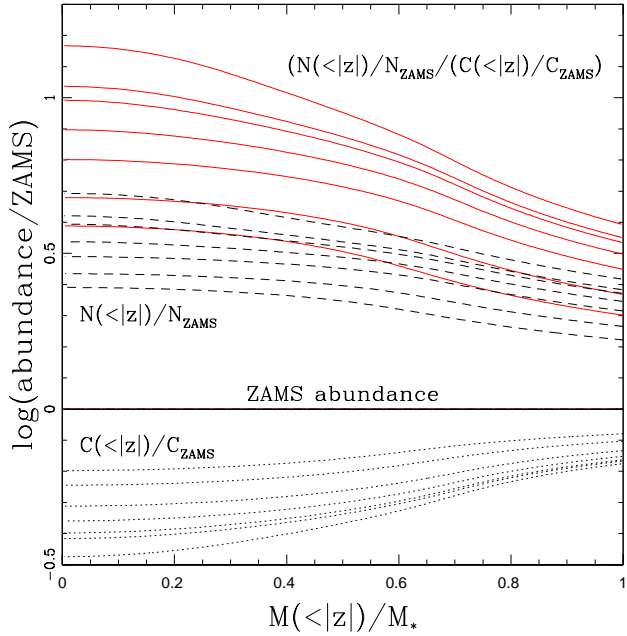}}
\caption{
  Enclosed carbon $C(<|a|)/C_{ZAMS}$ (black dotted), nitrogen $N(<|z|)/N_{ZAMS}$
 (black dashed) abundances and their ratio (red solid), all measured relative to
  the initial ZAMS values,  as a function of the enclosed
  mass fraction $M_*(<|z|)/M_*$ for the same times as in Figure~\protect\ref{fig:helium}.
  Already for the first profile distinguishable from the ZAMS abundances at $0.94$~Gyr,
  the nitrogen to carbon ratio is enhanced by over a factor of three.
  }
\label{fig:cno}
\end{figure}

\begin{figure}
\centerline{\includegraphics[width=3.5in]{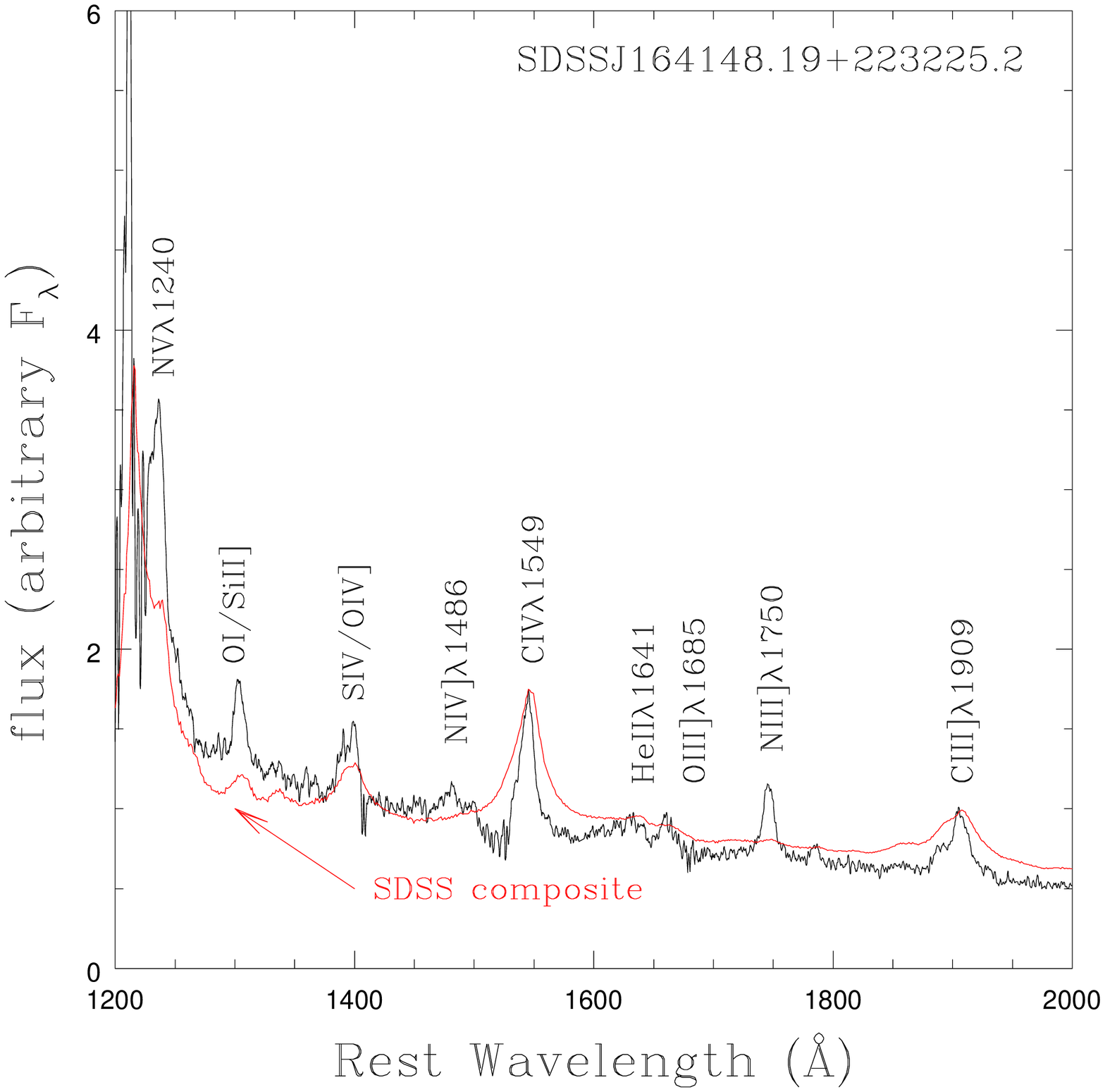}}
\caption{
  The spectrum of the nitrogen rich quasar SDSS~J164148.19$+$223225.2 at $z=2.51$
  (black) identified by \protect\cite{Batra2014} in comparison to the SDSS composite 
  quasar spectrum from \protect\cite{vandenberk2001} in red.
  }
\label{fig:nitrogen}
\end{figure}

\section{Abundance Anomalies in TDEs}

For our analyses we ran generic Solar metallicity ($X=0.70$, $Y=0.28$, $Z=0.02$)
{\tt MESA} (\citealt{Paxton2013}) models.
 In Figures~\ref{fig:mass0.5} to \ref{fig:mass2.0},
we show the evolution of the helium, carbon and nitrogen
abundances for $M_*=0.5M_\odot$, $1M_\odot$ and 
$2M_\odot$ in three regions: the entire star, outside
the inner edge of the hydrogen burning zone, and outside
the outer edge of the hydrogen burning zone.  The zone
edges are defined by the radius where the hydrogen 
energy generation rate exceeds $1$~erg~g$^{-1}$~s$^{-1}$.
These roughly represent the regions of the star that 
are likely to be disrupted.
On the main sequence, typically the star will be completely
disrupted (e.g. \citealt{Evans1989}, \citealt{Guillochon2013}).  
Evolved stars, however, are only likely to
be stripped into the hydrogen burning zone because of
the huge jump in density at its base.  The stripping
of the envelope may also require several pericentric
passages rather than occurring a single event (\citealt{Macleod2012}).

Figure~\ref{fig:mass0.5} shows that low mass stars simply
live too long to develop abundance anomalies given the age
of the universe.  Over its main sequence life time, an
$0.5 M_\odot$ star becomes significantly enhanced in 
helium and nitrogen and depleted in carbon, but 10~Gyr
represents less than 10\% of its overall lifetime
of $\sim 130$~Gyr.  Unlike more massive stars, the
carbon and nitrogen anomalies grow slowly because 
the p-p chain dominates the fusion rates.  {\it Producing an abundance anomaly 
in a TDE requires an $\gtorder 1 M_\odot$ star.}

Sun-like stars live $\simeq 12$~Gyr, which means that they
will be disrupted at a roughly random evolutionary phase for galaxies
with a constant star formation rate and at a late 
evolutionary phase for galaxies dominated by early bursts
of star formation.  Figure~\ref{fig:mass1.0} shows the
average abundance evolution with age for a
$1M_\odot$ star.  The general evolution is the same
until the late phases (note that Figure~\ref{fig:mass1.0}
covers 99\% of the stellar lifetime rather than the 90\%
shown in Figure~\ref{fig:mass0.5}).  On the main 
sequence, it develops a modest average enhancement of 
helium, a somewhat larger depletion of carbon, and
a still larger enhancement of nitrogen.  By the mid-point
of its main sequence evolution, the $\left[ N/C\right]$ 
abundance ratio has increased by a factor of three 
compared to its initial value, and as it leaves the 
main sequence, the enhancement is roughly a factor of four.
While the p-p chain still dominates the overall fusion rate,
CNO cycle reactions are sufficiently important to more rapidly alter
the carbon and nitrogen abundances.
During the MS evolution, the average oxygen abundance slowly
drops, but only by $\sim 10\%$ of the carbon depletion.
As the star begins hydrogen shell burning, the anomalies
outside the helium core begin to drop again.  However,
first dredge up also mixes them into the outer parts of
the star beyond the shell. Figure~\ref{fig:mass1.0} 
extends a little past the horizontal branch, where we
first see a rapid increase in helium followed by a 
rapid increase in carbon.

Figure~\ref{fig:mass2.0} shows the results for a $2M_\odot$
star.  The general pattern is similar to that for a $1M_\odot$
star except that the time scales are greatly accelerated since the
life time of the star is now only $1.2$~Gyr.  The average 
helium abundance slowly rises, while there is a rapid 
average depletion of carbon and enhancement of nitrogen.
Oxygen is again slightly depleted on the MS.  Because
the CNO cycle is much more important, the overall scale of
the carbon and nitrogen anomalies is significantly larger.
At the mid-point of its lifetime, the $\left[ N/C \right]$
ratio is increased by a factor of 6 on average, and the abundance
ratio continues to increase slowly until the star leaves
the main sequence.  First dredge up again mixes some of the
processed material into the outer regions of the star.  In
the later evolutionary phases, the regions from the hydrogen
burning shell outwards have little enhancement in helium and
smaller changes in $\left[ N/C \right]$.  As
stellar lifetimes become short compared to the typical time
scales of star formation, the evolutionary phase at the time
of disruption will be increasingly independent of the star
formation history. 

Figure~\ref{fig:cn} summarizes the behavior of $\left[ N/C \right]$
anomalies with stellar mass.  For stars with lifetimes 
$<10$~Gyr ($M_* \ltorder M_\odot$), we show the change in 
abundance at $10$~Gyr. For more massive, shorter lived stars,
we show the peak increase in $\left[ N/C \right]$ excluding 
the regions inside hydrogen burning shells (which generally means 
the values close to main sequence turn off). Lower mass
stars have negligible increases in $\left[ N/C \right]$
because the universe is too young and the p-p chain dominates
the reaction rates, while higher mass stars can
have nitrogen enhanced relative to carbon by an order of
magnitude due to the increasing importance of the CNO
cycle.

To make a rough estimate of the fraction of stars likely
to show anomalies, we need a scaling of the disruption
rate with stellar mass, a model for the IMF and a star
formation history.  If we follow \cite{Wang2004} or
\cite{Macleod2012}, the TDE rate scales as $R_*^{1/4} M_*^{-1/12} \simeq M_*^{1/6}$
if we roughly scale $R_* \propto M_*$ for the MS.  Higher
mass stars have lower densities and so have higher rates.
Stars closer to the main sequence turn off also have lower
densities than on the ZAMS, but we make no attempt to include
this effect.  For the IMF we follow \cite{Kroupa2001}
with $(dn/dM_*)_{IMF} \propto (M_*/M_b)^{-x}$ with $x=-1.3$ for
$0.08 M_\odot < M_* < M_b=0.5M_\odot$ and $x=-2.3$ for
$M_* > M_b$.  We show the rate per logarithmic interval
of mass, $M_* (dr/dM_*)$, where the rate per unit
mass, $dr/dM_*$, is proportional to the scaling of
the TDE rate with stellar mass, the IMF, and the number of
stars available at the mass given the star formation history.

Figure~\ref{fig:cn} also shows the scaling of TDE rates with 
stellar mass given these assumptions and three possible star formation histories: an
old burst of star formation, continuous star formation and
a recent burst of star formation.  The rates are all normalized
by the rate at $M_*=M_\odot$. 
For an old stellar population (``old burst'' in Figure~\ref{fig:cn})
we assume 2~Gyr of constant star formation starting 10~Gyr
in the past.  For this model, the rate is proportional to the
product of the IMF and the scaling of rates with stellar mass
for stars with main sequence lifetimes less than $\sim 8$~Gyr 
followed by an abrupt cut off because main sequence lifetimes 
$t_{MS}(M_*)$ are such a steep function of mass.  For $t_0=10$~Gyr
of continuous star formation, the number of stars at a given
mass scales as the IMF multiplied by $\hbox{min}(t_0,t_{MS})$, which produces a less
abrupt break near $M_*=M_\odot$.  Finally, we show the relative
rates for a recent 1~Gyr starburst (``young burst'').  
The combined effects of the dropping IMF and the rapid 
evolution of higher mass stars means that they contribute
little to the rate of TDEs independent of the star formation
history.   The slow evolution of low mass stars mean that they
contribute little to abundance anomalies.  As a result, nitrogen rich,
carbon poor TDEs should be sign posts for the disruption of
$1$-$2M_\odot$ stars.  For these various assumptions, the
rate of TDEs with $M_* \gtorder M_\odot$ is of order 7\% to 14\%
of the total rate.
However, when the fraction is low, the stars will tend to 
be closer to the main sequence turn off where the abundance
anomalies are large, and {\it vice versa} when the fraction is high.  
Hence, a reasonable order of magnitude estimate is that $\sim 10\%$
of TDEs will show strong abundance anomalies.

So far we have only examined the average abundances of the
material likely to form the debris of a TDE.  In practice,
the processed material is highly stratified and concentrated
towards the center with negligible changes outside the hydrogen
burning regions until first dredge up on the red giant
branch.  The disruption process sorts the debris by
orbital period (e.g. \citealt{Rees1988}), where the period is basically determined  
by the distance of the material $z$ along the direction
towards the black hole at pericenter. For a parabolic 
orbit, the binding energy of the debris is $-G M_{BH} z/R_p^2$, 
implying a semi-major axis of $a=R_p^2/2z$.  If we disrupt 
at $R_p=R_T=R_*(M_{BH}/M_*)^{1/3}$, then the orbital period is
\begin{eqnarray}
    P &= &{ \pi \over \sqrt{2} } \left( {M_{BH} \over M_* } \right)^{1/2}
        \left( { R_* \over z } \right)^{3/2} \left( { R_*^3 \over G M_*  }\right)^{1/2} \nonumber \\
      &= &0.11 \left( { R_* \over z } \right)^{3/2}~\hbox{years}
     \label{eqn:period}
\end{eqnarray}
for Solar units and $M_{BH}=10^6M_\odot$.  Hence, we
next examine abundance anomalies as a function of $|z|$, 
corresponding to planar slices through the star from 
the slice through the center at $z=0$ (with a nominally
infinite period for a parabolic orbit) to a slice just
grazing the surface at $|z|=R_*$.  While the distribution
in $|z|$ is symmetric, material with $z>0$ is bound and
can potentially be accreted, while material with $z<0$ is
unbound.  Hence, we
next examine abundance anomalies as a function of $|z|$, 
To minimize the number of cases, we will just
consider $M_*=1M_\odot$.  Higher mass stars show
qualitatively similar properties, and lower mass stars 
are not very interesting because they have too little time to evolve. 

Figure~\ref{fig:helium} shows the enclose helium mass 
faction, $Y(<|z|)$, as a function of the enclosed mass
fraction $M(<|z|)/M_*$ over the course of a $M_*=1M_\odot$
star's main sequence evolution.  The helium abundance in
the core steadily increases.  While the average abundance
grows by only $\simeq 35\%$ and the surface abundance is
little changed, a significant fraction of the debris mass
starts to be enhanced in helium by over 50\% after the MS 
lifetime, and by over 100\% at the end of the MS.
Keep in mind that these central values are for slices
through the star, so they are really averages of a 
slice through the core that is nearly 100\% helium 
with a slice through the envelope that still has the 
initial helium abundance.  

Figure~\ref{fig:cno} shows the changes in carbon,
nitrogen and their ratio relative to their initial
values as a function of the enclosed mass. 
Unlike the slow
increase in the helium abundance, the initial depletion
of carbon and enhancement of nitrogen is very rapid. Even
by an age of $\sim 1$~Gyr, half the mass of the star
has an average enhancement in $N/C$ of more than a factor
of three.  The rate of change then slows, but near the
end of the star's MS lifetime, the average enhancement in
the $N/C$ abundance ratio is roughly a factor of 10 for
half the mass of the star.

The one potential problem with exploiting these abundance
anomalies is that for a disruption at $R_p \simeq R_T$,
the orbital time scales for debris produced by the core
are long (years to decades), as also illustrated in 
Figure~\ref{fig:helium} based on Equation~\ref{eqn:period}.
The material accreted during the peak of the TDE light 
curve consist of material from the stellar surface layers
with the initial ZAMS abundances.  If having the processed
debris material contribute to the emission lines requires that
it complete an orbit and be processed through the resulting
radiation-hydrodynamic complexities, then we must use
disruptions at smaller pericentric radii, $R_p$.  Since the characteristic 
orbital time scales shorten $\propto (R_p/R_t)^3$, only modest 
reductions in $R_p$ rapidly accelerate the evolution. 
  
The second, and presently debated, possibility, is that photoionization
of the debris stream contributes directly to the observed emission
line spectrum.  The fundamental issue is the solid angle subtended
by the stream relative to the black hole, where the typical estimate
for a normal quasar is that the broad line clouds have a covering
fraction of order $f \simeq 10\%$
(e.g. \citealt{Peterson1997}).  The structure of the 
debris orbits leads to a roughly constant angular width in the
orbital plane as seen from the black hole (see, e.g., \citealt{Kochanek1994}).  
If the debris freely expands and maintains
a roughly fixed spread in orbital inclination, then the debris
stream can have a covering fraction of order a few percent and
it will contribute significantly to the observed emission line
spectrum, as in the models of \cite{Strubbe2009}.  If, on the
other hand, the focusing effects of self-gravity and tides
(see \citealt{Kochanek1994}) lead to the vertical scale height
expanding more slowly than the radius (roughly $r \sim t^{2/3}$), 
then the covering fraction of the debris stream is so small
that it is probably unimportant to the observed spectrum,
as in the models of \cite{Guillochon2014}.  An additional issue
is that the material with abundance anomalies will tend to lie
at the stream center rather than 
the surface unless there is mixing.  Since much of the stream is
optically thick to photo-ionizing radiation (\citealt{Kochanek1994}),
emission lines generated by photoionizing the surface layers will
show no evidence of the underlying abundance anomalies.

\section{Nitrogen Rich Quasars and TDEs}

The nitrogen rich quasars are a rare spectral class, representing roughly
1\% of SDSS quasars (\citealt{Bentz2004a}, \citealt{Bentz2004b}, \citealt{Dhanda2007}, \citealt{Jiang2008}).
\cite{Jiang2008} argue that the nitrogen rich quasars are similar in redshift, luminosity,
continuum slope, and black hole mass to most other quasars but have narrow carbon
lines and a much higher radio loud fraction. \cite{Batra2014} argue that the nitrogen
rich quasars are biased to lower black hole masses, and that the strong nitrogen
emission may require only high metallicity rather than nitrogen enhancement.  It
is difficult, however, to devise a stellar population that can produce enhanced
nitrogen abundances unless the absolute metallicity is very high
(see the discussion in \citealt{Batra2014}).  For illustration,
Figure~\ref{fig:nitrogen} shows the SDSS spectrum of the nitrogen
rich quasar SDSS~J164148.19$+$223225.2 identified by \cite{Batra2014} 
in comparison to the SDSS composite
quasar spectrum from \cite{vandenberk2001}.

We instead propose that the nitrogen rich quasars are in fact TDEs.  From 
\S2, it is clear that TDEs provide a natural source of nitrogen-rich, 
carbon-poor material --   the   
disruption of a single star easily provides
sufficient mass to pollute the broad line region (BLR) of a quasar,
since the mass in the BLR is only $\sim 10^{-3} M_\odot$ (e.g. \citealt{Peterson1997}).
The challenge is that, at least for Q0353$-$383,  the anomalous 
spectra can persist for decades (\citealt{Osmer1980}, \citealt{Baldwin2003}).
The time scale problem from \S2 is now reversed, in that we would
now like longer orbital time scales than years to decades.  This
can be easily achieved if the disrupted star is a giant rather
than a dwarf even for relatively low mass black holes ($10^6 M_\odot$,
see, e.g., \citealt{Macleod2012}), and then the time scale can be
even longer of higher mass black holes. 

However, as discussed
in \S2, disruptions of giants should be relatively rare, so
for our proposal to be plausible, the rate required to explain
the nitrogen rich quasars should be lower than the rate for
all evolved stars.  In 
\cite{Kochanek2015}, we estimate that the disruption rate of all evolved
stars at $z \sim 2$ is $\sim 5 \times 10^{-9}$~Mpc$^{-3}$~year$^{-1}$.
\cite{Batra2014}
found $N=43$ nitrogen rich quasars in the redshift range $2.29 < z < 3.61$ from 
the 5740~deg$^2$ of SDSS DR5 (\citealt{Adelman2007}), corresponding to a
survey volume of $V \simeq 89.4$~Gpc$^{3}$ ($\Omega=0.3$, $\Lambda=0.7$, $h=0.7$).
The mean redshift of the quasars is $\langle z \rangle = 2.80$.  If the
rest-frame lifetime of the event is $\Delta t = 10 \Delta t_{10}$~years, then
the required event rate is 
\begin{equation} 
        r = { N  \over V (1+\langle Z \rangle) \Delta t } \simeq 
          1.3 \Delta t_{10}^{-1} \times 10^{-11}~\hbox{Mpc}^{-3}~\hbox{year}^{-1}.
\end{equation}
Alternatively, \cite{Jiang2008} identified $N=293$ nitrogen rich quasars
in DR5 with $1.7 < z < 4.0$, corresponding to a survey volume of
$V \simeq 154$~Gpc$^{3}$, with $\langle z \rangle = 2.23$, yielding
a comparable rate estimate of
$ r \simeq 5.9 \Delta t_{10}^{-1} \times 10^{-11}$~Mpc$^{-3}$~year$^{-1}$.
These are low enough compared to the estimated TDE rate for evolved stars
in \cite{Kochanek2015} that there seems no difficulty in creating 
the nitrogen rich quasars using the very long lived TDE's associated
with the larger evolved stars.

A more general puzzle about abundances inferred from quasar BLRs
is that they appear to require very high metallicities, $Z \gtorder 5 Z_\odot$
(e.g., \citealt{Dietrich2003}, \citealt{Nagao2006}).  It is somewhat
challenging to reach these metallicities with global star
formation models (e.g., \citealt{Hamman1993}, \citealt{Friaca1998}, 
\citealt{Romano2002}), leading to an alternative picture of
star formation associated with the outer, self-gravitating 
parts of the accretion flow (e.g., \citealt{Collin1999}, \citealt{Wang2011}).
BLR metallicity estimates use the steady increase of nitrogen
relative to oxygen and carbon as the overall metallicity 
increases (\citealt{Hamman1993}). The N/C abundance
ratios of $M_* \gtorder M_\odot$ TDEs would naturally explain
the observed abundance ratios without the need for stellar populations
that are greatly super-Solar.  

The challenge is producing enough disrupted mass to reasonably
contaminate BLRs over long time scales.  For quasars accreting
near the Eddington limit, the accretion rate is of order
$\dot{M}_E \simeq 0.2 M_{BH7} M_\odot$/year where the black
hole mass is $M_{BH}=10^7 M_{BH7} M_\odot$ and assuming 
a radiative efficiency of 10\%.  A rough estimate of the
TDE rate for $M_*>M_\odot$ stars is $r\simeq 10^{-5}$/year
below $M_{BH} \simeq 10^7 M_\odot$, dropping to 
$r \simeq 10^{-6.5}$/year at $M_{BH} \simeq 10^9 M_\odot$ 
(see \citealt{Kochanek2015}).  The drop at higher black hole masses
is due to the increasing fraction of stars that pass through the
event horizon without disrupting.  Since the disrupted mass
is of order $M_\odot$, the mass accretion rates supplied
by TDEs, $\dot{M}_{TDE} \sim r M_\odot \sim 10^{-5}M_\odot$/year,
 are small compared to the average accretion rates of
active black holes.  This is simply a version of the point by 
\cite{Magorrian1999} that TDEs cannot be a major component
of quasar growth -- it is feasible to temporarily contaminate
the region around a quasar using TDEs as we propose for the
nitrogen rich quasars, but is it not feasible
to sustain the contamination in steady state.  

The one 
interesting loophole is the observation by \cite{Arcavi2014}
that a large fraction ($\sim 90\%$) of TDEs seem to be in post starburst
galaxies, which otherwise represent only $\sim 1\%$ of galaxies
(e.g., \citealt{Quintero2004}).  Assuming post starburst galaxies
represent recent mergers, and that mergers also are an important
trigger of quasar activity (e.g. \citealt{Hopkins2006}), then
the TDE rate when a quasar is active could be 100 times higher
than when a quasar is inactive.  Physically, the merger could
temporarily lead to very rapid diffusion of stars into low
angular momentum orbits or the presence of a binary black
hole could greatly increase the rates (see, e.g., \citealt{Li2015}).  This
would still not compete with the overall accretion rate, but
the overall rate at which TDEs would add mass to the region
near the black hole, $\dot{M}_{TDE} \sim 10^2 r M_\odot \sim 10^{-3}M_\odot$/year,
would no longer be negligible compared to the mass of the BLR.

\section{Discussion}

A natural consequence of stellar evolution is that the debris
in a TDE can show signs of nuclear processing.  Because TDEs
are likely dominated by lower mass stars ($M \ltorder 0.8M_\odot$) 
with very long evolutionary time scales, a large fraction 
should show no abundance anomalies.  However, TDEs of more
massive stars can have significantly enhanced helium and
nitrogen abundances and depleted carbon abundances.  The
helium anomalies grow quasi-linearly with the elapsed 
fraction of the star's main sequence life time, while the
carbon and nitrogen anomalies develop very quickly.  
We roughly estimate that $\sim 10\%$ of TDEs can show
significant abundance anomalies.  It is not feasible
to predict how this might be modified by 
observational selection effects.  

The average enhancement of helium is limited because it 
already represents a significant fraction of the stellar
mass at birth, and helium is likely sequestered in the 
dense helium core once the star starts to ascend the giant
branch.  However, at fixed orbital period, debris
including the core of a MS star can reach an average composition 
with roughly twice the initial helium mass fraction.  
Moreover, if the material at fixed period is not well-mixed,
it actually consists partly of material
with the initial helium abundance and partly of material that
is almost entirely helium.  Depending on the hydrodynamic
evolution of the debris and the details of the geometry,
the line emission observed by particular observers may be
dominated by material with radically different local
helium abundances even though the mean helium abundance is
only modestly increased.  

While not our focus, we should also note that
TDEs occur at the centers of galaxies that typically contain
stellar populations extending to higher metallicities than
Solar.  For example, our bulge contains stars of roughly twice
Solar metallicity (e.g., \citealt{Gonzalez2015}), so $X=0.64$, $Y=0.32$ and $Z=0.04$ for
an initial number fraction of $f=n(He)/(n(H)+n(He))=0.11$
instead of $f=0.09$.  A 25\% increase in the helium number fraction 
over Solar is neither huge nor negligible for addressing the line ratio 
problem discussed in \S1.  Furthermore,
there are arguments that metal rich bulge stars are also helium
enriched beyond this standard scaling (see, e.g., \citealt{Nataf2015}).
As discussed in \S3, such super-Solar stars will also have
larger ratios of $N/C$, although stars as metal rich as
those invoked to explain quasar metallicities are very
rare (or non-existent) in the Galactic bulge.

These TDEs should also show abundance anomalies in carbon, which is
depleted, and nitrogen, which is enhanced. Oxygen also tends
to be slightly depleted but by amounts that are probably 
unobservable.  The changes in the carbon and nitrogen  
abundances occur very quickly, with the changes occurring
in under 10\% of the MS life time for the $M_* \gtorder M_\odot$
stars that will show anomalies.  Half the mass of a
Sun-like star shows a change in the $N/C$ abundance 
ratio of over a factor of three almost immediately, 
growing to almost an order of magnitude by the main 
sequence turn off.  More massive stars with stronger
CNO cycles show still larger anomalies.  However, the
decline in both the IMF and stellar life times with
stellar mass means that TDEs with carbon/nitrogen anomalies 
should almost all be $\sim 1M_\odot$ to $2M_\odot$ 
stars, almost independent of the assumed star formation 
history.  Lower mass stars take too long to evolve, and
higher mass stars are too rare.

For characterizing the abundance anomalies we used either
the overall averages or averages at a fixed post-disruption
period.  For observational signatures, this may underestimate
the importance of the processed material.  As
long as the disrupted material is not well-mixed, it
will tend to have some memory of its initial density because
its expansion and contraction will be roughly self-similar
(e.g., \citealt{Kochanek1994}). Thus, at fixed orbital period,
higher density regions of the star will tend to be higher
density regions of the debris.   If the medium is optically thin
and the density is low enough to avoid collisional de-excitation,
then the strength of both recombination lines and collisionally
excited lines are proportional to density squared rather than
the linear weighting used in \S2.
This means that for debris with any
``memory'' of its initial density, our 
results in \S2 could be underestimates of the observational
effects of the abundance anomalies.

Considerable progress is being made in hydrodynamic
simulations of TDES (e.g., \citealt{Macleod2012}, \citealt{Dai2013}, \citealt{Guillochon2014},
\citealt{Shiokawa2015}, \citealt{Bonnerot2016}), although
full radiation hydrodynamic simulations are needed
given the likely importance of radiation to the flow 
properties (e.g., \citealt{Loeb1997}, \citealt{Strubbe2009},
\citealt{Miller2015}, \citealt{Strubbe2015}, \citealt{Metzger2015}). 
For probing the role of abundance anomalies, simulations need to
track the distribution of debris as a function of abundance and/or
initial radius within the star.  First, this would address
the question of mixing and residual correlations of abundance
with density. Second, it would clarify the structure and thermal
state of the debris streams.  Third, for material that has returned to pericenter, it is important to
determine if the debris ends up ordered inside out (i.e. long
period material ends up at small radii due to hydrodynamics)
or outside in (i.e. remains ordered by the post-disruption orbital
binding energy).  

For lower mass black holes, it is possible for a star to pass
sufficiently close to the black hole to disrupt the far denser
helium core of an evolved star. For simplicity we did not
explore the details of such cases, since the event rate will
be very low compared to the disruptions of main sequence
stars or the envelopes of evolved stars.  Obviously, such
disruptions would allow still higher helium fractions 
and greater changes in CNO abundances.  Anomalies in
heavier elements, with the possible exception of s-process elements 
in the disruption
of an AGB star, are unlikely to ever be observed because
it requires the disruption of a rare massive star in the
very (very!) short carbon burning phase.  

A long standing puzzle in quasar spectra has been the small
population ($ \sim 1\%$) of nitrogen rich quasars (e.g.
\citealt{Bentz2004a}, \citealt{Bentz2004b}, \cite{Dhanda2007},
\citealt{Jiang2008}, \citealt{Batra2014}).  There
are no natural mechanisms for evolving stellar populations
to produce a nitrogen rich interstellar medium other
than stellar populations with extremely high metallicities
and even this may be problematic (see the discussion in \cite{Batra2014}).   
TDEs can naturally supply a significant mass
of nitrogen rich/carbon poor material, well in excess of 
the total mass required for a typical broad line region,
and so could provide an explanation for the nitrogen rich
quasars.  Making the anomaly long lived (decades) would
require the disruption of giant stars in order to have
long time scales for the TDE, but the numbers of nitrogen
rich quasars appear to be low enough to be compatible with
the low rates of such events.  This hypothesis does
require, however, that the BLR emission of the nitrogen
rich quasars should evolve with time, so it would be
interesting to survey the existing samples of nitrogen
rich quasars for spectral changes.

It is difficult to use TDEs to explain the super-Solar
metallicities estimated for quasars in general 
(e.g., \cite{Dietrich2003}, \cite{Nagao2006}), basically
because the average mass accretion rate from TDEs
is small compared to the average accretion rate of
black holes (\citealt{Magorrian1999}).  If, however,
the observation by \cite{Arcavi2014} that a large
fraction of TDEs are associated with post-starburst
galaxies continues to hold, then the TDE rate during
periods of quasar activity might also 
be enhanced by a factor of $\sim 10^2$.  
While still a small fraction
of the mean accretion rate, the TDEs would supply a
significant amount of material compared to that 
contained in the BLR.    

The known TDEs are all at low redshift, where optical 
spectra will show signatures of anomalies in helium
but not carbon and nitrogen.  Searching for carbon and
nitrogen anomalies requires ultraviolet spectra to probe
the wavelength regions shown in Figure~\ref{fig:nitrogen}.  This is
feasible with the Hubble Space Telescope, and recent
observations by \cite{Cenko2015} show that the
TDE ASASSN-14li (\citealt{Holoien2015}) has a UV
spectrum very similar to that of a nitrogen rich 
quasar, with strong NV, NIV$]$  and NIII$]$ lines
and relatively weaker CIV and CIII$]$ lines, as well
as relatively strong HeII emission compared to 
normal quasars.  This both bolsters our hypothesis
for the origin of the nitrogen rich quasars and
strongly suggests that ASASSN-14li was due to the
disruption of a $1$-$2M_\odot$ star.  Because of the
mapping between abundance and the original location
of the material in the disrupted star, monitoring 
the evolution of such spectra should explore the 
evolving hydrodynamics of TDEs.

\section*{Acknowledgments}
We thank R. Pogge for reminding us about the peculiar nitrogen rich quasars,
J. Johnson, M. Pinsonneault and J. Tayar for answering many questions about 
stellar evolution and composition, T. Thompson for discussions and B.
Shappee for comments.  
CSK is supported by NSF grants AST-1515876 and AST-1515927.

\end{document}